\documentclass[12pt]{iopart}
\usepackage{epsf}
\begin{document}

\title{Anisotropy in the helicity modulus of a
3D XY-model: application to YBCO}

\author{Bo\v zidar Mitrovi\' c
\footnote[3]{To whom correspondence should be addressed
(mitrovic@newton.physics.brocku.ca)}, Shyamal K. Bose, and Kirill
Samokhin}

\address{Physics Department, Brock University, St. Catharines,
Ontario, Canada L2S~3A1}

\begin{abstract}
We present a Monte Carlo study of the helicity moduli of an
anisotropic classical three-dimensional (3D) XY-model of
YBCO in superconducting state. It is found that both the
$ab$-plane and the $c$-axis helicity moduli, which are
proportional to the inverse square of the corresponding
magnetic field penetration depth, vary linearly with
temperature at low temperatures. The result for the
$c$-axis helicity modulus is in disagreement with the
experiments on high quality samples of YBCO. Thus we
conclude that purely classical phase fluctuations of
the superconducting order parameter cannot account for
the observed $c$-axis electrodynamics  of YBCO.
\end{abstract}



\maketitle

\section{Introduction}

Emery and Kivelson presented [1,2] very strong physical arguments
that in systems with low superfluid density $n_{s}$ (e.g.,
high-T$_c$ copper oxide superconductors and the organic
superconductors) the fluctuations in phase of the superconducting
order parameter play a significant role. In essence, the
superfluid density is proportional to the helicity modulus of a
superconductor [3], which measures the stiffness of the
superconductor with respect to twists in phase of the order
parameter; hence a low value of $n_{s}$ implies a low value of the
maximum possible phase ordering temperature $T_{\phi}$. If
$T_{\phi}$ is comparable to the measured physical superconducting
temperature T$_c$, the observed superconducting properties may be
very different from what is predicted by the mean-field
BCS/Eliashberg theory. They further argued that in the case of
poor screening, as indicated by a low conductivity, the Coulomb
interaction suppresses the local Cooper pair density fluctuations
$\Delta n_{s}$, which in turn enhances the phase fluctuations
$\Delta \phi$ ($\Delta n_{s}\Delta \phi\geq$ 1/2). Thus, the phase
fluctuations might influence the superconducting properties over a
wide range of temperatures below and above T$_c$. Indeed,
measurements of the $ab$-plane magnetic field penetration depth
$\lambda(T)$ ($\lambda^{-2}\propto n_{s}$) on
YBa$_{2}$Cu$_{3}$O$_{6.95}$ [4] gave unambiguous evidence for
three-dimensional (3D) XY critical scaling behavior in a
temperature interval of about 10 K below T$_c$=92.74 K. Roddick
and Stroud [5] were the first to point out that classical phase
fluctuations in a nodeless order parameter could produce a
low-temperature $\lambda(T)$ which varies linearly with $T$ in an
isotropic three-dimensional superconductor. They also showed that
this dependence persists when combined effects associated with
Coulomb energy of local charge density fluctuations and Ohmic
dissipation are included in the model. This important result
showed that the observed linear temperature dependence of the
$ab$-plane $\lambda^{-2}$ in YBCO at low T [6] does not
necessarily originate from quasiparticle excitations near the
lines of the gap nodes at the Fermi surface. The same idea was
advanced independently by Emery and Kivelson [1], who also argued
that a large $ab$-plane conductivity in YBCO implies that the
phase fluctuations are predominantly {\em classical} down to low
temperature. More recently, Emery and Kivelson [7] argued that
very extensive angle-resolved photoemission spectroscopy
experiments found no evidence of quasiparticle excitations near
the nodes of the gap even just below T$_c$, and therefore the
existence of nodes is not responsible for the observed linear
$T$-dependence of the $ab$-plane $\lambda^{-2}$ in YBCO at low
temperatures.\\

If, indeed, the phase fluctuations in the order parameter are
responsible for the observed linear temperature dependence of the
$ab$-plane superfluid density at low $T$ [1,5,7], then the
experiments [8] on $c$-axis electrodynamics in YBCO are puzzling
and have to be explained.  Namely, it was found that the $c$-axis
penetration depth $\lambda_{c}(T)$ never has the linear
temperature dependence observed in $ab$-plane. The best fit to the
data up to about 40 K gives $\Delta\lambda_{c}(T)\propto T^{2.1}$.
If the phase fluctuations are predominantly classical down to low
temperature one would expect the superfluid density to be linear
at low $T$ in {\em all} directions. Indeed, that is exactly what
was obtained by Roddick and Stroud [5] in a simple model for an
anisotropic superconductor using a variational self-consistent
phase phonon (SCPP) approximation. Here we present a detailed
Monte Carlo (MC) study of helicity moduli for an anisotropic 3D
XY-model which incorporates the bilayer structure of
YBa$_{2}$Cu$_{3}$O$_{7}$. As expected, we find the helicity moduli
to decrease linearly at low temperatures in all directions. Thus
the classical anisotropic 3D XY-model cannot account for the
observed temperature dependence of the magnetic field penetration
depth of YBa$_{2}$Cu$_{3}$O$_{7}$ along the $c$-axis.

The rest of the paper is organized as follows. In section 2 we
discuss our model and some details of the MC calculation. Section
3 contains our numerical results for the helicity moduli and our
conclusions.
\section{The model and the simulation}

We consider an anisotropic 3D XY-model for a system with bilayer
structure described by the Hamiltonian
\begin{eqnarray}
H & = & H_{ab}+H_{c}\>,\\
H_{ab} & = & \sum_{l,s,\langle i,j\rangle}\left[-J_{1}
\cos(\phi_{i,l,s}-\phi_{j,l,s})
-J_{2}\cos(2(\phi_{i,l,s}-\phi_{j,l,s}))\right]\>,\\
H_{c} & = & \sum_{l,i}\left[- J_{\perp}
\cos(\phi_{i,l,2}-\phi_{i,l,1}) -
J_{\perp}'\cos(\phi_{i,l+1,1}-\phi_{i,l,2})\right]\>.
\end{eqnarray}
Here, the sum over $l$ runs over a stack of bilayers, the sum over
$s=1,2$ runs over two layers in a given bilayer, $\langle
i,j\rangle$ denotes the nearest neighbors within a single layer,
the sum over $i$ runs over the sites in a given layer and
$\phi_{i,l,s}$ is the phase of the order parameter on site $i$ of
the layer $s$ in the bilayer $l$. The ab-plane Josephson couplings
with constants $J_{1}$ and $J_{2}$ correspond to the transfer of
one and two Cooper pairs [9], $J_{\perp}$ is the Josephson
coupling constant between two layers within a given bilayer, and
$J_{\perp}'$ is the Josephson coupling constant between layers in
two adjacent bilayers. This Hamiltonian is obtained from the
coarse-grained Ginzburg-Landau model for YBCO [10] by ignoring the
temperature dependence of the amplitude of the order parameter.

The computation of the helicity modulus along direction
perpendicular to the bilayers requires care as the separation
between the layers in a bilayer, $c_{b}$, is, in general,
different from the separation between the bilayers, $c'$. If a
{\em uniform} vector potential ${\bf A}$ is applied its effect on
the Hamiltonian is to shift the phase difference between points
${\bf r}_{1}$ and ${\bf r}_{2}$ by $A_{1,2}=2\pi {\bf A}\cdot({\bf
r}_{2}-{\bf r}_{1}) /\Phi_{0}$, where $\Phi_{0}=hc/2e$ is the flux
quantum: $ \phi_{{\bf r}_{1}}-\phi_{{\bf r}_{2}}\longrightarrow
\phi_{{\bf r}_{1}}-\phi_{{\bf r}_{2}} +A_{1,2}$. The momentum
conjugate to $\phi_{i,l,s}$ is the charge $n_{i,l,s}$ (in the
units of $2e$) and its rate of change, given by $-\partial
H/\partial\phi_{i,l,s}$, is the total current flowing into the
site $(i,l,s)$. The resulting expression can be used to define the
current along each bond connected to site $(i,l,s)$. In
equilibrium $\langle\dot{n}_{i,l,s}\rangle$ = 0 and one obtains
the {\em average} current conservation at each site $(i,l,s)$. For
a uniform ${\bf A}$ perpendicular to the layers, the average
currents parallel to the bilayers vanish by symmetry and one has
\begin{equation}
\langle J_{\perp}\sin(\phi_{i,l,2}-\phi_{i,l,1}+\frac{2\pi}{\Phi_{0}}Ac_{b})\rangle =
\langle J_{\perp}'\sin(\phi_{i,l+1,1}-\phi_{i,l,2}+\frac{2\pi}{\Phi_{0}}Ac')\rangle\>,
\end{equation}
where the angular brackets denote the statistical average for the
Hamiltonian $H(A)$: $\langle\cdots\rangle$ = $\int\Pi
d\phi_{i,l,s}\cdots\exp(-\beta H(A))/Z(A)$, $\beta$ =
$1/(k_{B}T)$, $Z(A)$ = $\int\Pi d\phi_{i,l,s}\exp(-\beta H(A))$.
The expression on the left hand side of Eq.~(4) gives the average
current from the site $i$ in $s$ = 1 layer of the bilayer $l$ into
the site $i$ of $s$ = 2 layer of the same bilayer. The expression
on the right hand side of (4) gives the average current from the
latter site into the site $i$ of $s$ = 1 layer in the adjacent
layer $l$+1.

Since the average currents along bonds parallel to ${\bf A}$ are
equal, one can calculate the $z$-axis helicity modulus (the
$z$-axis is perpendicular to the layers, i.e.\ it is along the
crystallographic $c$-axis) by computing the derivative of any one
of those average currents with respect to $A$ at $A$ = 0. For
computational purposes using the Monte Carlo method we evaluate
the average helicity modulus over all bonds parallel to ${\bf A}$
\begin{eqnarray}
    \fl\gamma_{zz} = \frac{2\pi}{\Phi_{0}}\left\{\frac{1}{N^3}\sum_{l,i}\left[J_{\perp}c_{b}
    \langle\cos(\phi_{i,l,2}-\phi_{i,l,1})\rangle+J_{\perp}'{c'}\langle\cos(\phi_{i,l+1,1}-\phi_{i,l,2})\rangle
    \right]\right. \nonumber \\
    -\frac{1}{k_{B}T}\left\langle\frac{1}{N^3}\sum_{l,i}\left[J_{\perp}\sin(\phi_{i,l,2}-\phi_{i,l,1})+
    J_{\perp}'\sin(\phi_{i,l+1,1}-\phi_{i,l,2})\right]\right. \nonumber \\
    \left. \times\sum_{l,i}\left[J_{\perp}c_{b}\sin(\phi_{i,l,2}-\phi_{i,l,1})+
    J_{\perp}'c'\sin(\phi_{i,l+1,1}-\phi_{i,l,2})\right]\right\rangle \nonumber \\
    +\frac{1}{k_{B}T}\frac{1}{N^3}\sum_{l,i}\left[J_{\perp}\langle\sin(\phi_{i,l,2}-\phi_{i,l,1})\rangle
    +J_{\perp}'\langle\sin(\phi_{i,l+1,1}-\phi_{i,l,2})\rangle\right] \nonumber \\
    \left. \times\sum_{l,i}\left[J_{\perp}c_{b}\langle\sin(\phi_{i,l,2}-\phi_{i,l,1})\rangle+
    J_{\perp}'c'\langle\sin(\phi_{i,l+1,1}-\phi_{i,l,2})\rangle\right]\right\}\>.
\end{eqnarray}
Here, $N^3$ is the number of bonds parallel to ${\bf A}$. We note
that computing $\gamma_{zz}$ as $\partial^{2}F/\partial
A_{z}^{2}$, where $F$ is the Helmholtz free energy, which is valid
(up to a multiplicative constant) for orthorhombic lattices
without a basis, would give a wrong result in our case. The
in-plane helicity modulus $\gamma_{xx}$ along $x$- (i.\ e.\ $a$-)
direction is derived in analogous way.

The elementary excitations of the XY-model are spin-waves and
vortices [3], but in the zero temperature limit only the
spin-waves are statistically relevant. It is useful to have the
analytic expressions for $\gamma_{zz}$ and $\gamma_{xx}$ obtained
from the spin-wave expansion in the zero temperature limit as a
check on the Monte Carlo results at low temperatures. We find (in
the units of $(2\pi/\Phi_{0})2N$, where $N$ is the number of
tetragonal unit cells, and taking $J_{2}=0$ for the sake of
simplicity)
\begin{equation}
    \fl\gamma_{zz}(T)=
    \frac{1}{2}\left[(J_{\perp}c_{b}+J_{\perp}'c')-
    (J_{\perp}-J_{\perp}')(J_{\perp}c_{b}-J_{\perp}'c')/(J_{\perp}+J_{\perp}')\right] -\alpha_{z}k_{b}T\>,
\end{equation}
where
\begin{eqnarray}
    \alpha_{z} &=& \frac{1}{4N}\sum_{\bf k}\left[(J_{\perp}c_{b}+J_{\perp}'c')
    \left(\frac{1}{\omega_1({\bf k})}+\frac{1}{\omega_{2}({\bf k})}\right)\right. \nonumber\\
    && -\left. (J_{\perp}c_{b}
    \kappa({\bf k})+J_{\perp}'c'\kappa'({\bf k}))\left(\frac{1}{\omega_{1}({\bf k})}-
    \frac{1}{\omega_{2}({\bf k})}\right)\right]\>,
\end{eqnarray}
\begin{equation}
\kappa({\bf k})=(J_{\perp}+J_{\perp}'\cos(k_{z}c))/\sqrt{(J_{\perp}+J_{\perp}')^{2}-
4J_{\perp}J_{\perp}'\sin^{2}(k_{z}c/2)}\>,
\end{equation}
\begin{equation}
\kappa'({\bf k})=(J_{\perp}\cos(k_{z}c)+J_{\perp}')/\sqrt{(J_{\perp}+J_{\perp}')^{2}-
4J_{\perp}J_{\perp}'\sin^{2}(k_{z}c/2)}\>,
\end{equation}
\begin{eqnarray}
\omega_{1,2}({\bf k})= &2J_{1}\left(\sin^{2}\frac{k_{x}a}{2}+\sin^{2}\frac{k_{y}a}{2}\right)
+ (J_{\perp}+J_{\perp}')/2\nonumber\\
&\mp \frac{1}{2}\sqrt{(J_{\perp}+J_{\perp}')^{2}-4J_{\perp}J_{\perp}'\sin^{2}(k_{z}c/2)}\>.
\end{eqnarray}
In Eq.~(7) the sum over ${\bf k}$ runs over $(k_{z}\geq0)$-half of
the Brillouin zone $-\pi\leq k_{x}a,k_{y}a,k_{z}c\leq\pi$ and
$c=c_{b}+c'$. The analytic expression for $\gamma_{xx}$ is much
simpler due to homogeneity of in-plane couplings and bond lengths
\begin{equation}
\gamma_{xx}= J_{1}a\left[1 -\frac{k_{B}T}{2N}\sum_{\bf k}\sin^{2}\frac{k_{x}a}{2}
\left(\frac{1}{\omega_{1}({\bf k})}+\frac{1}{\omega_{2}({\bf k})}\right)\right]\>.
\end{equation}

To study the helicity moduli of the model described by the
Hamiltonian (1-3) we have used Monte Carlo simulations based on
the Metropolis algorithm [12]. In our finite-lattice MC
simulations we have used periodic boundary conditions on lattices
with 10$\times$10$\times$10, 20$\times$20$\times$20 and
30$\times$30$\times$30 sites. For a given set of parameters
$J_{1}$, $J_{2}$, $J_{\perp}$, $c_{b}$, $J_{\perp}'$, $c'$, the
simulation would start at a low temperature with all phases
aligned. The first 250 000 MC steps per site (sps) were thrown
away followed by seven MC links of 250 000 MC sps each. At each
temperature the range over which the phase angle was varied [13]
was adjusted to ensure the MC acceptance rate of about 50\%. The
errors were calculated by breaking up each link into 5 blocks of
50 000 MC sps, then calculating the average values for each of 35
blocks and finally taking the standard deviation $\sigma$ of these
35 average values as an estimate of the error. The final
configuration of the phase angles at a given temperature was used
as a starting configuration for the next higher temperature. The
cumulant analysis [13] was performed using the results for the
three lattice sizes to determine the value of the transition
temperature $T_{c}$.

\section{Numerical results and conclusions}

We have performed MC simulations for different choices of
parameters and since they give qualitatively similar results we
present here only the data for a single set of parameters $J_{1}$
= 1, $J_{2}$ = 0, $J_{\perp}$ = 0.1, $J_{\perp}'$ = 0.04, $c_{b}$
= 0.4, $c'$ =1. The values of $c_{b}$ and $c'$ were chosen so that
their ratio reflects the structure of
YBa$_{2}$Cu$_{3}$O$_{7-\delta}$ [14]. The value of Josephson
coupling between the layers of a bilayer is taken to be one order
of magnitude smaller than the in-plane Josephson coupling and the
value of Josephson coupling between bilayers $J_{\perp}'$ was
chosen such that $J_{\perp}c_{b}$ = $J_{\perp}'c'$. Since the
helicity modulus is proportional to inverse square of the magnetic
field penetration depth $\lambda(T)$ [5], we present our results
for the helicity moduli in Fig.~1 as
$\lambda^{2}(0)/\lambda^{2}(T)$ as a function of $T/T_{c}$. From
the cumulant analysis of the numerical results for the three
lattice sizes in in Fig.~1 we obtained $k_{B}T_{c}$ = 1.265 (in
the units of $J_{1}$). This value should be compared to 0.89 --
0.95 for a two-dimensional square lattice [15-17] and  2.2 for a
three-dimensional cubic lattice [18-20] (which we have also
reproduced with our code) in the units of the coupling constant.

\begin{figure}
\begin{center}
\epsfbox{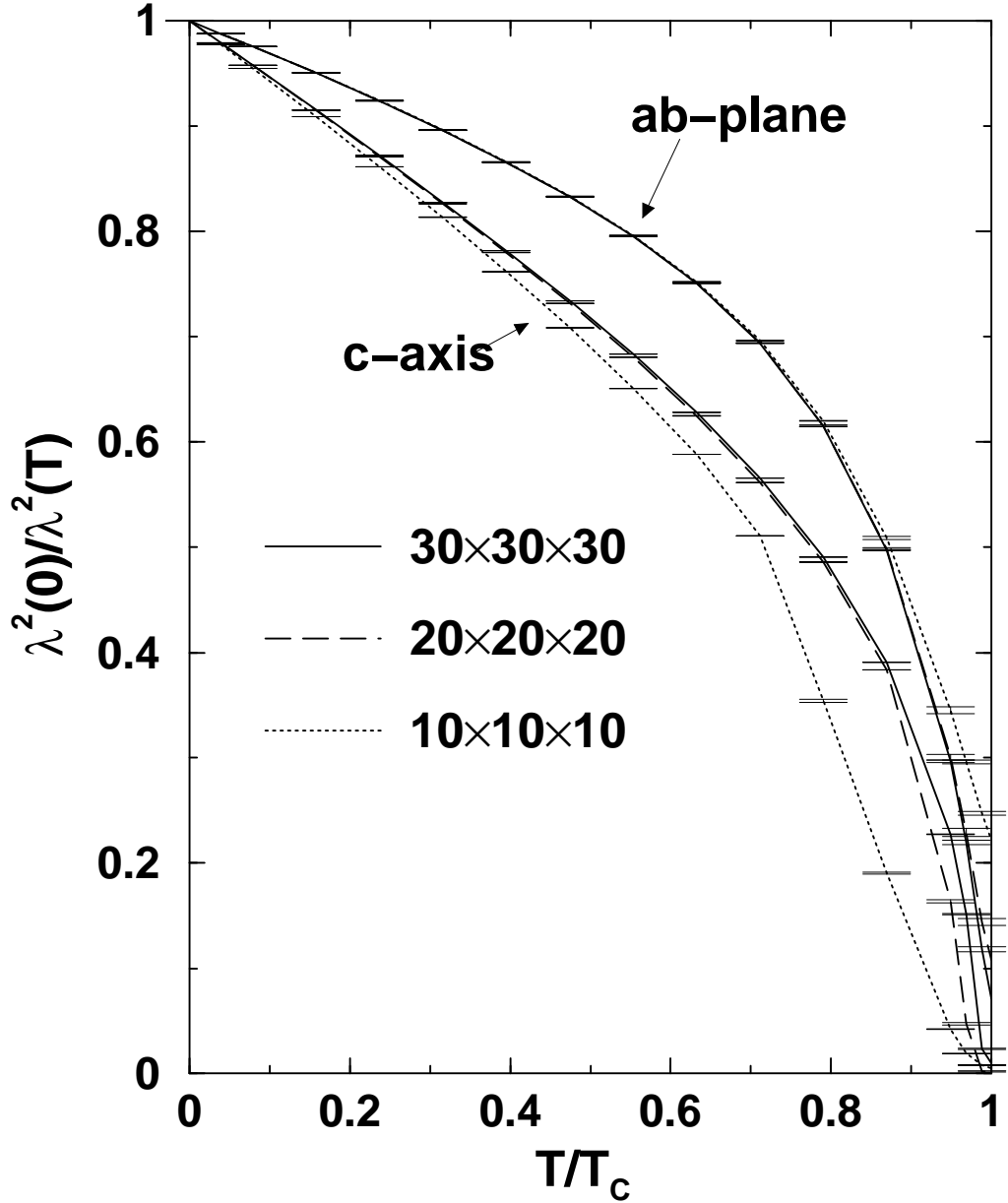}
\end{center}
\caption{\label{label}The temperature dependence of helicity
moduli for three different lattice sizes.}
\end{figure}

As anticipated, $\lambda^{2}(0)/\lambda^{2}(T)$ is decaying
linearly at low T both along the $ab$-plane and along the $c$-axis
with the slopes of -0.2366 $\pm$ 0.0001 and -0.415 $\pm$ 0.001,
respectively, as determined by the linear fits to the
low-temperature data ($T\leq$0.01$T_{c}$). These values should be
compared with -0.16667 $\pm$ 0.00005 which we found for our
isotropic 30$\times$30$\times$30 lattice with the periodic
boundary conditions. We note that the spin-wave expansion formulae
(5-11) give the slopes of 0.2367 along $ab$-plane, -0.404 along
$c$-axis for our anisotropic 3D XY-model, and -1/(6$J_{1}$) =
-0.1666$\dot{\mbox{6}}/J_{1}$ for the isotropic 3D XY-model, in
excellent agreement with our Monte Carlo result. Note that our
results for the $ab$-plane helicity modulus are somewhat lower
than the spin-wave result 0.25/$J_{1}$ for two-dimensional square
lattice, which is easily understood from Eqs. (10-11) when
$J_{\perp}$ and $J_{\perp}'$ are one to two orders of magnitude
smaller than $J_{1}$.

In terms of the superfluid density fraction
$\rho_{s}=\lambda^{2}(0)/\lambda^{2}(T)$ as a function of
$t=T/T_{c}$ the slopes are -0.30 along $ab$-plane, -0.52 along
$c$-axis for our anisotropic 3D XY-model, and -0.37 for the
isotropic 3D XY-model (compare with the value -0.4 reported in
[7]).

Our calculations clearly demonstrate that {\em classical} phase
fluctuations cannot explain the observed nonlinear temperature
dependence of the $c$-axis magnetic field penetration depth for
high-$T_{c}$ samples of YBCO [8].

\ack
This work has been supported in part by the Natural Sciences and Engineering
Research Council of Canada.
\end{document}